\documentclass[aps,prc,nofootinbib,amsmath,amssymb]{revtex4}
\usepackage{eurosym}
\usepackage{amsmath,amssymb}
\usepackage{graphicx}
\usepackage{dcolumn}
\usepackage{bm}
\usepackage{color}
\usepackage{multirow}
\usepackage{float}
\usepackage{array}
\usepackage{dcolumn}
\usepackage{booktabs}
\usepackage[detect-all]{siunitx}
\usepackage{makecell}
\usepackage{xcolor}
\usepackage{yfonts}
\usepackage[utf8x]{inputenc}

\graphicspath{ {/Users/Nastia/Desktop/research_dis/papers_writing/xenes/nov_5} }

\newcolumntype{P}[1]{>{\centering\arraybackslash}p{#1}}
\begin{document}

\title{Electron-impact ionization and ionic fragmentation of O$_{2}$ from
threshold to 120 eV energy range.}
\author{ R. A. Lomsadze$^{1}$, M. R. Gochitashvili$^{1}$, R. Ya. Kezerashvili$^{2,3}$, and M. Schulz$^{4}$  }
\affiliation{ \mbox{$^{1}$Department of Exact and Natural Sciences, Tbilisi State
University, 0179 Tbilisi, Georgia} \\
\mbox{$^{2}$Physics Department, New York City College
of Technology, The City University of New York,} \\
Brooklyn, NY 11201, USA \\
\mbox{$^{3}$The Graduate School and University Center, The
City University of New York, \\
New York, NY 10016, USA }\\
\mbox{$^{4}$Missouri University of Science and Technology, Rolla, MO 65409, USA}
}
\date{\today}

\begin{abstract}
We study the electron-impact induced ionization of O$_{2}$ from
threshold to 120 eV using the electron spectroscopy method. Our approach is
simple in concept and embodies the ion source with a collision chamber and a mass spectrometer with a quadruple filter as a selector for
the product ions. The combination of these two devices
makes it possible to unequivocally collect all energetic fragment
ions formed in ionization and dissociative processes and to detect them with known efficiency.
The ion source allows to vary and tune the electron-impact ionization energy and the target-gas pressure.
We demonstrate that for obtaining reliable results of cross sections for
inelastic processes and determining mechanisms for the formation of O$%
^{+}$($^{4}S,^2{D},^2{P}$) ions, it is crucial to control the electron-impact energy for production of ion and the pressure in the ion source.
A comparison of our results with other experimental and theoretical data shows
good agreement and proves the validity of our approach.

\end{abstract}


\maketitle




\section{Introduction}

The N$_{2}$ and O$_{2}$ molecules are the main constituents of the atmospheres and are also important in other gaseous media.
The importance of obtaining an understanding of their response to electron impact, which occur in the
aurora and discharges due to the interaction with electron component of the solar wind, has prompted a large number of experimental and theoretical investigations
of these species. Total electron scattering cross sections are well known as are those for ionization,
elastic scattering, rotational and vibrational excitation, dissociation, and for the excitation to several of the
electronic states in these species. Such comprehensive coverage is required
by our expanding need to observe and predict the
complex solar-terrestrial interactions that comprise space
weather. For space weather applications, N$_{2}$ and O$_{2}$ provide the primary source
of satellite drag that impacts orbit determination and tracking \cite{15}.

Electron-impact ionization of oxygen and nitrogen atoms and molecules is
of fundamental importance in atmospheric science, atmospheric
and man-made plasmas processes, and mass spectrometry and has a very long history. An extensive review of the literature reveals, that these processes have been studied very extensively over many decades since 1930s \cite{McDaniel89,Lindsay2002}. The oxygen atoms and ions, molecules and molecular ions are a simple species of great interest for interstellar
medium and planetary atmospheres, not only for the Earth, but for other planets and their moons like Venus, Saturn, Europa, Ganymede, Titan \cite{1RL, 2RL, Hall1995, 4RL, Luna2003, Europa2005, 3RL, 5RL}.
The atmosphere of Europa was first detected by Hall
et al. \cite{Hall1995} using Hubble Space Telescope observations and the existence of oxygen has been confirmed by several studies, see, for example, Ref. \cite{5RL}. Gas
phase ionization processes of atmospheric molecules induced by photons,
electrons or by excited metastable neutral species play a key role in
numerous phenomena occurring in low energy ionized plasmas. Ionic spaces are
present in the upper atmosphere of planets and govern the chemistry of
ionospheres. Among all, atomic and molecular ions have been detected in
comet tails.

Other important aspects of electron-impact ionization of oxygen atoms are the production of the metastable states O$^{+}$$(^{2}P$) and O$^{+}$$(^{2}D$).  It is well known that metastable states of oxygen atoms play significant roles in auroral emissions observed in the Earth's polar
regions as well as in the cometary coma and airglows. The metastable states O$^{+}$$(^{2}P$) and O$^{+}$$(^{2}D$) play a
crucial role in the atmospheric chemistry at mesospheric and thermosphere
altitudes. A theoretical study of electron impact ionization of initially metastable
states of nitrogen and oxygen atoms and their relevance to auroral emissions is presented in Ref. \cite{Pandya2014}. The O$^{+}$($^{2}P$) state undergoes quenching by reacting with different
constituents of the atmosphere and a study of the various
quenching mechanisms by which the O$^{+}$($^{2}P$) ion is lost in
the atmosphere is important \cite{15, Singh2009}. Oxygen is also one of the typical impurities in almost all
laboratory plasmas \cite{6RL}. Among the many inelastic processes involving
singly charged O$^{+}$ ions, the excitation,
ionization and charge exchange are relevant to the low-temperature edge
plasma region of current thermonuclear fusion devices \cite{6RL, Plasma}.
Electron impact ionization is the most commonly adopted procedure for laboratory-generated
plasma and for aerospace applications \cite{Shang2018}.

O$^{+}$ is the dominant ion in the F region
of the ionosphere and both metastable species of O$^{+}$$(^{2}P$) and O$^{+}$$(^{2}D$) have been detected there. Metastable O$^{+}$ ions are important species in a variety
of situations ranging from electrical discharges to astrophysical phenomena
(see, e.g. \cite{7RL,8RL}). The metastable ions interact with neutrals in the extended
regions of the atmosphere. The charge exchange produces fast neutrals that are unaffected by the local fields and can directly penetrate into the atmosphere
making collisions with the atmospheric neutrals causing heating, collisional ejection of atoms
and molecules.

It is imperative to study different inelastic processes induced by the ground and metastable states O$^{+}$ ions with atoms and molecules.
One important reason for this is the fact that the cross section
is strongly dependent on the initial electronic state of the O$^{+}$ ions. For example, the total cross section for the electron capture by the metastable O$^{+}$$(^{2}D,^{2}P$) ions in collisions with He atoms at   keV energies is similar to or even greater that that for the ground state O$^{+}$$(^{4}S$) \cite{28}. The experimental results reported in Ref. \cite{21RLRKHe} showed that the metastable state cross sections are of the same order of magnitude as those for the ground state.  Also the finding showed that the metastable state exhibits a significantly higher
cross section compared to the ground state ions in the lower scale of their energy, while the
cross sections tend to be of the same magnitude as the collision energy reaches 5 keV \cite{Lindsay1997}. Therefore, the laboratory studies of different processes induced by collision of the ground and metastable states of O$^{+}$ with atoms and molecules requires to control the beam of O$^{+}$ ions.
The goal of this work is to study production of singly charged oxygen ions
by electron impact and to use them as a projectile to explore
various inelastic processes on atoms and molecules.

In this paper we study an electron impact ionization and ionic fragmentation of O$_{2}$ in the energy range from
threshold to 120 eV. The
paper is organized in the following way. In Sec. II we discuss the composition of O$^{+}$ ion beam.  An experimental
apparatus which allows one to perform measurements by means of electron
spectroscopy method is presented in Sec. III. Here we introduce our approach and
procedure for cross section measurements and its determination. We report
measurements with
the incident beam of O$^{+}$ in the ground and metastable states and discuss
results in Sec. IV. Finally, in Sec. V we summarize our
investigations and present conclusions.

\section{Composition of the O$^{+}$ ion beam}

\begin{figure}[t]
\centering
\includegraphics[width=15cm]{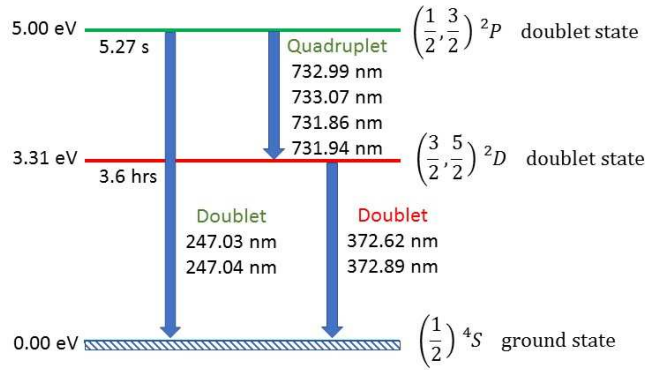}
\caption{(Color online) Low energy level diagram for O$^{+}$ ions. Not in scale}
\label{Fig11}
\end{figure}

One important motivation to study inelastic processes by O$^{+}$ ions is the
fact that the magnitude of the cross section is strongly dependent on the initial
electronic state of O$^{+}$ ions \cite{3, 4, 5}. Despite the fact that the reactions with O$^{+}$ incident ions have received
considerable attention and several experimental studies were
conducted (see e.g. \cite{14RL, 27} there still exists a large
amount of doubt the magnitude of cross section for elementary processes.
It turned out that
the oxygen ion beam, in addition to O$^{+}$($^{4}S$) ground state ions, can contain a
substantial fraction of two low-lying O$^{+}$($^{2}D$) and O$^{+}$($^{2}P$)
metastable ions, with approximate radiative lifetimes of 3.6 h and 5.3 s,
respectively \cite{6}. For convenience in Fig. \ref{Fig11} a simplified low energy
level diagram of atomic oxygen ions is given based on data from Refs. \cite{RK1spector, 15}.
Considering the spin-orbit splitting of the levels in Fig. \ref{Fig11}, there
are three radiative paths:\ the O$^{+}$($^{2}P$) and O$^{+}$($^{2}D$) transitions to the ground
state O$^{+}$($^{4}S$)  produce doublets ($\lambda =247.03$ nm and\ $%
\lambda =247.04$ nm) and ($\lambda =372.62$ nm and\ $\lambda =372.89$ nm), respectively, both in
the ultraviolet region and the transition O$^{+}$($^{2}P$) to the O$^{+}$($^{2}D
$) state produces a quadruplet ($\lambda =731.86$ nm, $\lambda =731.94$, and
$\lambda =732.99$ nm,  $\lambda =733.07$ nm) in the visual  spectrum. Based
on the transition probabilities the radiative lifetime of the O$^{+}$($^{2}P$%
) state is 5.27 sec, with 78\% radiating into the O$^{+}$($^{2}D$) \cite%
{Zeippen,15}.

It is well established that collisions of O$^{+}$ ions with atoms and molecules have in
many cases large cross sections and thus higher reaction rates than
corresponding collisions with the ground state species. Therefore, the existence of metastable excited states ions in the
primary ion beam, can sometimes significantly influence the observed cross
sections. The investigation of charge transfers processes for O$^{+}$ ions
in collisions with H$_{2}$ molecules for the O$^{+}$($^{4}S,^{2}D,^{2}P)$$-$H$_{2}$ \cite{21RLRK} and He atoms for  O$^{+}$($^{4}S,^{2}D,^{2}P)$$-$He \cite{28, 21RLRKHe} collision systems has shown that metastable
state ions indeed enhance the cross sections, sometimes by an order of
magnitude relative to that for the ground state O$^{+}$($^4{}$S) \cite{28}.

The large cross section for reactions involving O$^{+}$ incident ions in
metastable states and small cross section for reactions of O$^{+}$ ions
being in the ground state is explained by the energy defect of the considered process
(energy difference before and after reactions). For example, the charge
transfer reaction O$^{+}$($^{2}D$) -- H$_{2}$($X^{1}\Sigma _{g}^{+},v^{\prime \prime }=0$)
is nearly resonant
(energy defect is 0.15 eV). Hence the magnitude of the cross section obtained for excited
O$^{+}$($^{2}D$) ions is an order of magnitude larger than the one for the ground O$^{+}$($^{4}S)
$ state ions \cite{22}.


There have been many attempts related to the control of metastables in ion source
and to perform research with known ratio of metastable states to ground states
\cite{5,15,7,11,21RLRK,20RL,14,24RL}. Over 30 years ago it was pointed in Ref. \cite{5} that this ratio is related to the pressure in
the ion source and the impact-electron energy producing the ion. The ratio of metastable states to ground states was checked by
careful control of the energy of the ionizing electrons in the primary ion
source in Ref. \cite{25RL}. It means, when electron energy is below
the threshold for excited-state formation, no excited states can be formed
and the resultant beam will be composed entirely of ground- state ions. As
the ionizing electron energy increases the metastable ions will appear in
the beam. For example, ionization of O$_{2}$($X^{3}\Sigma _{g}^{-},v=0$)
molecules by
18.9 eV electrons results in the formation of ground state O$^{+}$($^{4}S$) which is
6.84 eV above the 12.06 eV ionization potential of  O$_{2}$($X^{3}\Sigma _{g}^{-},v=0$) \cite{25RL}.
The energy difference between the ground O%
$^{+}$($^{4}S$) and first excited O$^{+}$($^{2}D$) state is 3.32 \cite{15}.
Thus, the minimum electron energy required to produce excited O$^{+}$($%
^{2}D$) ions from O$_{2}$($X^{3}\Sigma _{g}^{-},v=0$) is 22 eV.

The relative abundances was determined for the O$^{+}$($^{4}S$), O$^{+}$($%
^{2}D$) and O$^{+}$($^{2}P$) ions for 100 eV electron impact ionization of O$%
_{2}$ in Ref. \cite{30}. Their finding is 0.55, 0.25 and 0.20,
respectively. It was shown (see e.g. \cite{32}) that the use of
high-pressure ion sources can lead to significant loss of metastable ions
due to collisional quenching mechanisms. A relatively low O$^{+}$($^{2}D$)
and O$^{+}$($^{2}P$) metastable ion abundance is observed in Ref. \cite%
{33} when the target pressure, on the order of 10$^{-1}$ Torr, is applied in
the ion source. The fractional abundance for different states of O$^{+}$ ion
beam formed during ionization of O$_{2}$ at the ionization electron energy 100 eV, is considered in Ref. \cite{29RL}.
This result indicated that 70\% of O$^{+}$ ions are in the ground $^{4}S
$ state and 30\% are in the $^{2}D$ state. Two-component types of attenuation for O$^{+}$ beams in N$%
_{2}$ gas are also found in \cite{29RL}. They ascribed this to
considerable (30\%) admixtures of metastable O$^{+}$ ion in the beam, which
could be varied by using different ionizing electron energies in the source.
In Ref. \cite{35} the authors found an appropriate range of an ion source pressure to have a certain percentage ratio for the projectile degree of excitation.
The quenching rate of O$^{+}$($^{2}P$) by various atomic and molecule spaces
have been evaluated in \cite{15}. Their findings agree well with
the measurements obtained by \cite{32RL} and a factor of 8 lower than
that obtained in Ref. \cite{33RL}. The composition of the ion beam with
the abundance about 65\% of ground state $^{4}S$ and about 35\% that of
metastable species $^{2}D$ is obtained in \cite{27}. The differences in the results for cross sections are mainly
due to different metastable fractions within the ion beams reported by various groups.

\section{EXPERIMENTAL APPARATUS}

The experimental study of the production of oxygen
ions in the ground and metastable states is performed
in three steps. First, to investigate the formation of O$^{+}$ ions we use
electron impact to produce charged ions and monitor the production by using electron spectroscopy. To
calibrate and check the mass transmission and electron energy we measure the electron induced
ionization on well studied targets such as CH$_{4}$ molecules \cite{44, CH4, methane2014} and Ar atoms \cite{Crowe1972, Straub95, Lindsay2002, Ar2005, Ar2006, Ren2012, Campeanu2019}. Second, we study the ion
production in $e + $O$_{2}$ collision as a function of the incident electron
energy for different pressure conditions of the molecular oxygen in the ion
source.

\begin{figure}
\begin{tabular}{cc}

\textit{(a)} &  \textit{(b)}\\
\
  \includegraphics[width=90mm]{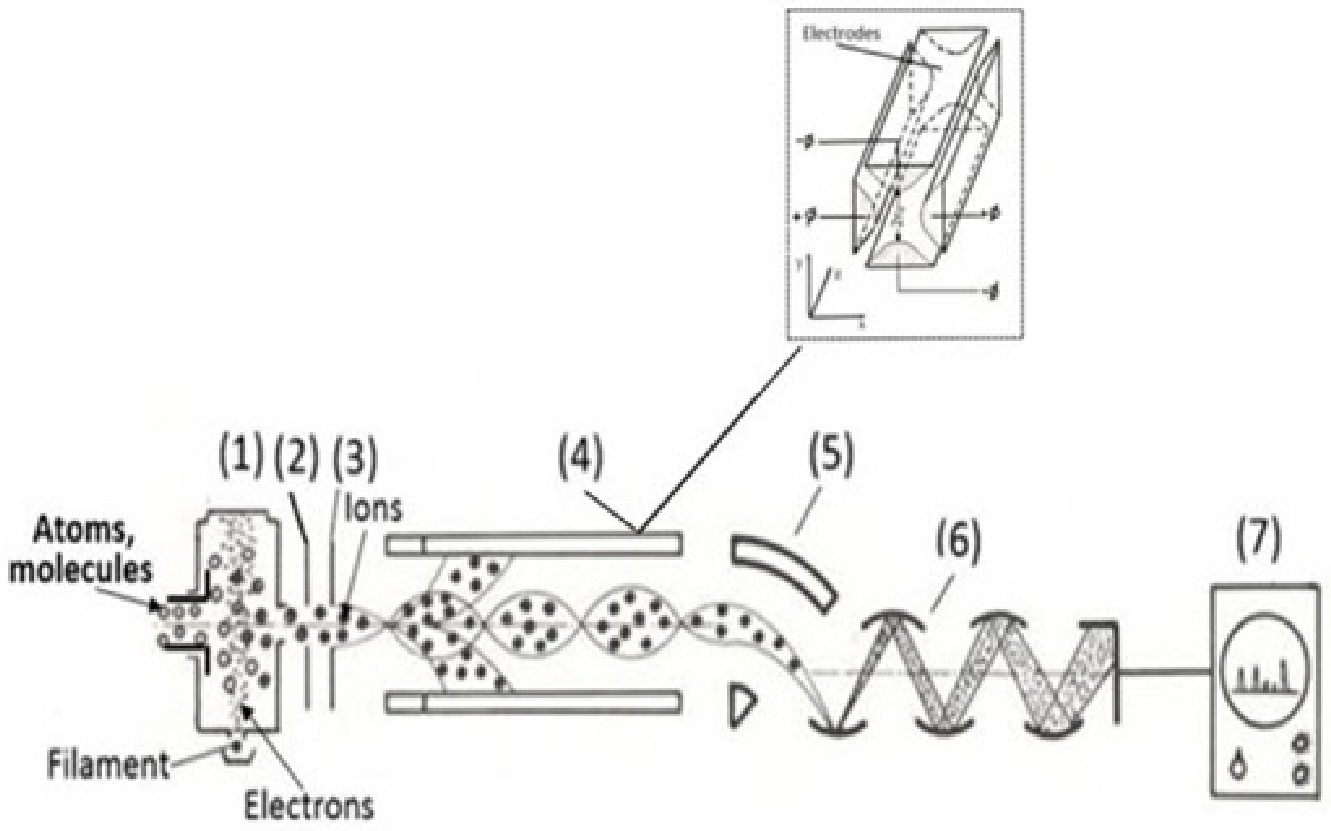} &   \includegraphics[width=70mm]{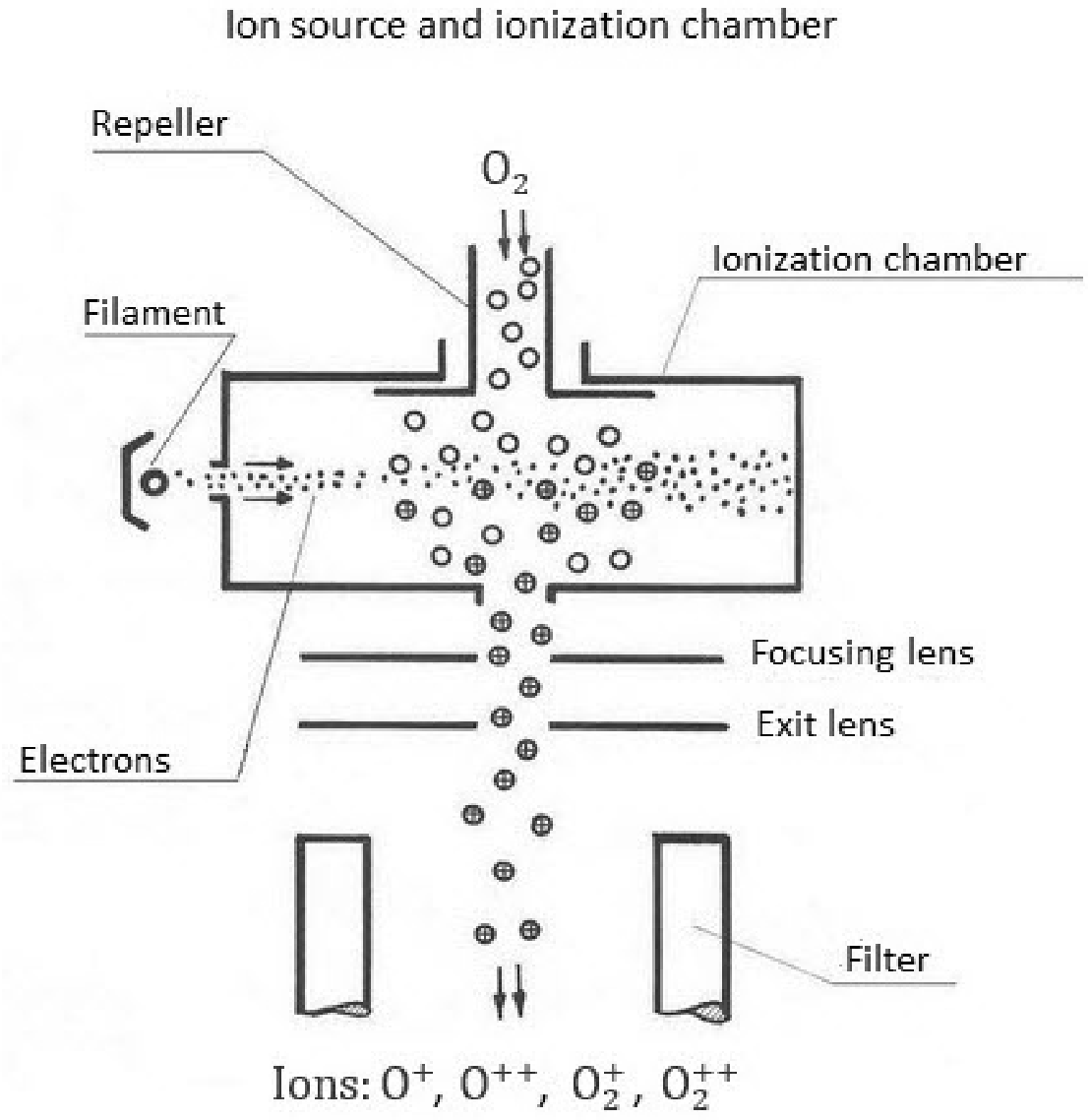} \\[6pt]

\end{tabular}
\caption{(Color online) $(a)$ Schematic diagram of the experimental setup: 1 -
ion source and ionization chamber, 2 - extracting lens, 3 - focusing lens, 4 -
quadruple filter, 5 - deflector, 6 - electron multiplier, 7 - acquisition
system. The inset at the upper right panel in Fig. \ref{Fig1}$a$ shows the device for a quadruple
filter. $(b)$ Enlarged view of the ion source and ionization chamber. The ion source allows for varying and tuning the electron-impact ionization energy and the target-gas pressure.
The image is rotated clockwise by a 90$^{0}$ relative to its position in Fig \ref{Fig1}$(a)$. Figures are not to scale.}
\label{Fig1}
\end{figure}

Measurements carried out with the electron impact source are performed by a
mass-spectrometric (MS) device, a schematic view of which is shown in Fig. %
\ref{Fig1}. The main MS assembly consists of (in sequential order): an
electron impact ionization ion source (1); extracting and focusing lenses
(2) and (3), respectively; a quadruple filter (4); deflector plates (5); an
electron multiplier (6); a data acquisition system (7). The apparatus is
partitioned into three chambers which are separately evacuated by two
mechanical and two diffusion pumps. The pressures are measured by two separate MKS Baratron gauges and are kept at about 10$^{-7}$ Torr.

At low electron energies, in the ionization chamber shown in Fig. \ref{Fig1},
the following dominant ionization processes of
the oxygen molecule with liberation of valence electrons occur:
\begin{align}
e+\text{O}_{2}& \rightarrow \text{O}_{2}^{+}+2e, \\
\text{ \ \ \ \ \ }& \Downarrow  \\
& \text{O}_{2}^{+}\text{(}X^{2}\Pi _{g}\text{)}+2e, \\
& \text{O}_{2}^{+}\text{(}a^{4}\Pi _{u}\text{)}+2e, \\
& \text{O}_{2}^{+}\text{(}A^{2}\Pi _{u}\text{)}+2e, \\
& \text{O}_{2}^{+}\text{(}b^{4}\Sigma _{g}^{-}\text{)}+2e; \\
& \rightarrow \text{O}_{2}^{++}+3e+...; \\
& \rightarrow \text{O}^{+}+2e+... \\
& \Downarrow  \\
& \text{O}^{+}\text{(}^{4}S\text{)}+2e+...; \\
& \text{O}^{+}\text{(}^{2}P\text{)}+2e+... \\
& \Downarrow  \\
& \text{O}^{+}\text{(}^{4}S\text{)}+2e+\gamma _{247}..., \\
& \text{O}^{+}\text{(}^{2}D\text{)}+2e+\gamma _{732}...; \\
& \text{O}^{+}\text{(}^{2}D\text{)}+2e+... \\
& \Downarrow  \\
& \text{O}^{+}\text{(}^{4}S\text{)}+2e+\gamma _{373}...,
\end{align}
where the ionization processes are accompanied by a release of an electron
from the 1$\pi _{g}$, 1$\pi _{u}$ and 3$\sigma _{g}$ states, respectively, for the ionized molecule O$^{+}_{2}$. Also the electron-impact dissociation of O$_{2}$ at electron energies between 13.5 and 198.5 eV form two oxygen atoms with production of O($^{1}D$) and  O($^{3}P$) fragments following electron impact excitation and ionization was reported in Ref. \cite{Cosby}. There are also many different processes related to the double ionization of oxygen atoms and molecules, as well as decays of O$_{2}^{+}$ into O$^{+}$ ions ground and metastable states, which we do not listed.
Basically, the target gas ions are generated in the
ionization region of the ion source by low energy electron impact from threshold to 120 eV. The electron
beam is produced by thermionic emission from a filament. The electrons are
than accelerated by the potential applied between the filament and the
ionization chamber. This potential defines the energy of the ionizing
electrons. The absolute energy of the electron beam has been checked by
measuring ionization potentials of rare gas atoms and diatomic molecules.
The electron energy is monitored during each experiment by a measurement of
the O$_{2}$ ionization potential and O$^{+}$ appearance potential as well.
The electron current is of order 0.2 mA. The produced O$^{+}$ ions inside the
ionization chamber are extracted and focused by appropriate ion optics. The
potential of the ionization chamber is adjustable and determines the energy
of the ions entering the quadruple filter. A quadruple filter shown in the
inset in Fig. \ref{Fig1}$a$ consists of an assembly of four parallel
electrodes with a hyperbolic cross section. Each pair of opposite electrodes
is electronically connected. The following potential is applied to one pair
of electrodes
\begin{equation}
F(t)=U+V\cos2ft,
\end{equation}
where $U$ is the DC potential, while $V$ and $f$ are the amplitude value and
frequency of an AC potential, respectively. An opposite potential $F(t)$ is
applied to the other pair of electrodes. Both DC and AC electrical fields
are directed perpendicular to the $z$ axis and produce lateral oscillations
of any ion entering the device parallel to the $z$ axis.

The motion of the O$^{+}$ ion in the filter is described by a set of
differential equations, known as the Mathieu equations, which can be
resolved into two types of movement -- stable or unstable -- according to
the value of parameters $U$, $V$, $f$, $r_{0}$ and $m/q$. Here $2r_{0}$ is
the distance between two opposite electrodes, and $m$ and $q$ are,
respectively, the mass and charge of the incoming ion. In practice, ion
selection is achieved by variations of the $U$ and $V$ potentials. For an
appropriate value of these two parameters, the ions injected into the
quadruple filter are separated, and only the ions with a value of $m/q$
ratio in a certain bandwidth will be transmitted by the filter. All the
other ions have unstable trajectories and are captured on the electrodes.
For a mass spectrometer the capacity to separate particles of similar mass
known as \textquotedblleft a resolving power\textquotedblright\ is defined
as the ratio $R=\frac{m/q}{d(m/q)}$, where $d(m/q)$ is the bandwidth of the
mass to charge $m/q$ ratios for which ions achieve stable trajectories. The
theoretical resolution of a quadruple filter depends on the value of the
ratio $V/U$: a decrease of this ratio increases the sensitivity. An amplitude
scan between zero and the maximum value if the DC and radio frequency voltages
applied to the electrodes yields a transmission by the filter of ions of
increasing $m/q$, with constant resolution. The entire source
assembly is heated to prevent cold areas, which would act as condensation
traps for the analyzed compounds. A cross section profile of the bars is
hyperbolic, as it is shown in the inset in Fig. \ref{Fig1}$a$. A careful
choice of design parameters such as the relation between the distance and
the diameter of the bars gives a good approximation of an ideal quadruple
field in the vicinity of the $z$ axis of the filter. The whole assembly is
bakeable for degassing. Ions selection is achieved by a manual or automatic
scanning of the DC and AC voltages. The amplitudes of the $U$ and $V$ potentials
are modulated by a saw tooth signal and the whole spectrum (from mass 2 to
mass 1000) can be explored in one second. The ion detector is an electron multiplier consisting of a chain of 21
copper-beryllium dynodes. The overall gain is of the order of 10%
$^{6}$ and it is controlled by a $3$ keV regulated power supply fed to the
dynodes through a chain of resistors. The electron multiplier is mounted
off-axis to reduce the noise produced by photons emitted from the source
filament and by soft X rays generated by electrons bombardment of the
rods of the quadruple filter. The output signal of the electron multiplier
is fed either into an analogy measurement signal or into a digital data
acquisition system for a further treatment by computer.

\begin{figure}[t]
\centering
\includegraphics[width=10.0cm]{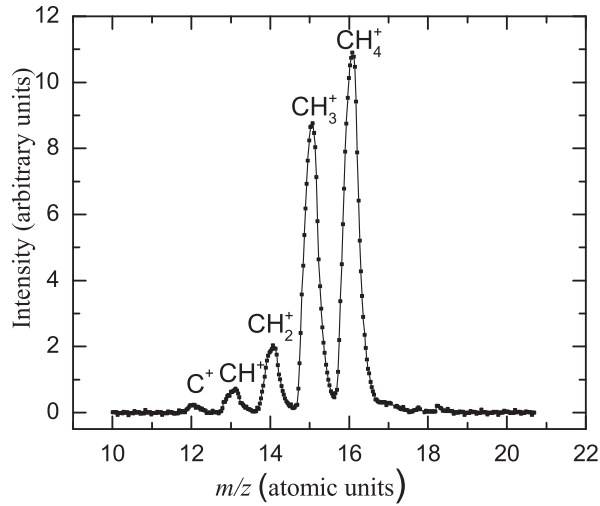}
\caption{(Color online) A typical ion fragment's mass spectrum of methane
for an incident electron energy of 115 eV. Species detected are as labeled
on the figure.}
\label{Fig2}
\end{figure}

\section{Results and discussion}

When performing
measurements for electron impact it is important to control the energy of
electrons and detect all positively charged ions with equal efficiency,
regardless of their mass, charge or initial kinetic energy. Let us first
present results of our investigation on electron induced ionization, dissociative
ionization and ionic fragmentation of hydrocarbon CH$_{4}$ molecules and Ar
atoms in the energy range from the ionization threshold up to 120 eV. In
particular, we are conducting studies for the following reactions:
\begin{align}
e+\text{CH}_{4} & \rightarrow\text{CH}_{4}^{+}+2e \\
& \rightarrow\text{CH}_{3}^{+}+2e+... \\
& \rightarrow\text{CH}_{2}^{+}+2e+... \\
& \rightarrow\text{CH}^{+}+2e+...
\end{align}
\begin{align}
e+\text{Ar} & \rightarrow\text{Ar}^{+}+2e \\
& \rightarrow\text{Ar}^{++}+3e
\end{align}
as well as for reactions (1), (7) and (8). The motivation of these studies is to check and calibrate the mass
transmission of the quadrupole spectrometer and to provide data of the
appearance energy (AE) and the partial ionization (PI) cross section, which
may serve as input parameters for the optical spectroscopy measurements
of excitation processes with oxygen ion beam. The measurements were
performed using a quadruple mass spectrometer at a precisely selected
electron energy. The mass dependence of transmission of the quadrupole spectrometer is
performed by measuring the mass spectra of CH$_{4}$ molecule for reactions $%
(19)-(22)$ and by comparing these results with the cross sections for CH$_{i}^{+}$
($i=1,2,3,4)$ fragment ions \cite{44}. Our measurement of a typical ion
fragment's mass spectrum for methane molecules is shown in Fig. \ref{Fig2},
which is in good agreement with previous results \cite{44}. A number of
well-resolved mass peaks are detected in the mass range $10-20$ amu and are
assigned to the corresponding ionic fragments. As can be seen, our mass
resolution is somewhat limited, so that fragments differing only by one mass
unit are not fully resolved. Therefore, the results for separate dissociative products of hydrogen atomic and molecular ions  with masses of 1 and 2 are not shown in Fig. \ref{Fig2}. It is found, that the transmission for the ratios (C$%
^{+}$/CH$_{4}$) and (CH$_{4}^{+}$/CH$_{4}$) with masses of 12 and 16, respectively is constant. This is the mass region where the majority of the relative efficiency of the oxygen fragments, O$^{++}$,  (O$^{+}-$O$_{2}^{++}$) and O$_{2}^{+}$ appear. It was assumed that the mass of oxygen fragment ions does not change appreciably for masses outside of 12 to 16 mass region and the transmission for the masses of 8, 16 and 32 amu  was believed to equal to that of 12-16 amu.

\begin{figure}[b]
\centering
\includegraphics[width=7.0cm]{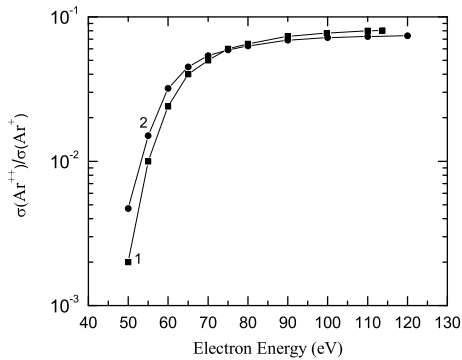}
\caption{(Color online) Dependence of the ratio of double to single
ionization cross sections of Ar atom in the $e+$Ar collision on the electron
energy. Curves: 1 - present data; 2 - theoretical prediction \protect\cite%
{Lindsay2002}.}
\label{Fig3}
\end{figure}

Significant efforts are put into the determination
of appearance energies of ionic fragments. For this reason the electron
energy was calibrated against the known AE of Ar$^{+}$ (15.7 eV) to within $%
\pm\ 0.25$ eV. \ As a result, the ionic fragment cross section curves
obtained in the $50-120$ eV range are used to confirm the correctness of this
approach. We check this approach by performing measurements of the
ionization cross sections for the processes (23) and (24). The results of the
ratio for (Ar$^{++}$/Ar$^{+}$) double to single ionization cross section of
the Ar atom as a function of impacting electron energy are presented in Fig. %
\ref{Fig3}. On the same figure the theoretical results are presented for
comparison. The experimental errors for these data, including the accuracy
of the present normalizing procedure, have been estimated at about 10\%. As
can be seen from Fig. \ref{Fig3}, our cross section data show a general
agreement with the calculations from Ref. \cite{Lindsay2002} within 7\%. The close
agreement between these data sets demonstrates the expected efficient
collection of ion fragments in our apparatus.

\begin{figure}[t]
\centering
\includegraphics[width=10.0cm]{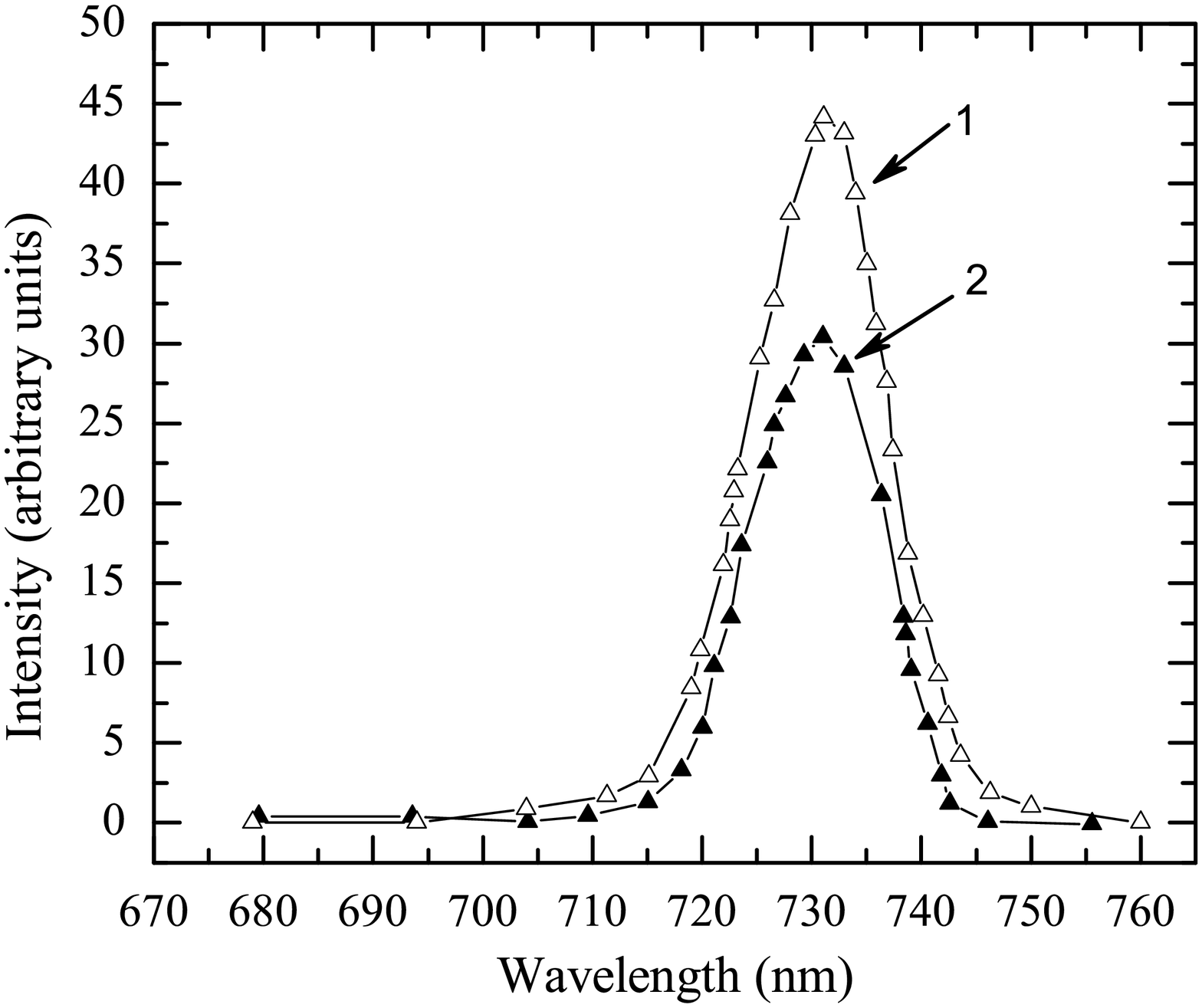}
\caption{(Color online) The typical spectra of Ar atomic line ($\protect%
\lambda=731.1$ nm) for O$^{+}-$Ar collision system at fixed ($E=2.5$ keV)
energy of O$^{+}$ colliding ions and for two different pressure in
high-frequency ion source. Curves: 1 - for the pressure $P=1.5\times10^{-2}$
Torr; 2 - for the pressure $P=1\times10^{-1}$ Torr. }
\label{Fig9}
\end{figure}

\begin{figure}[t]
\centering
\includegraphics[width=10.0cm]{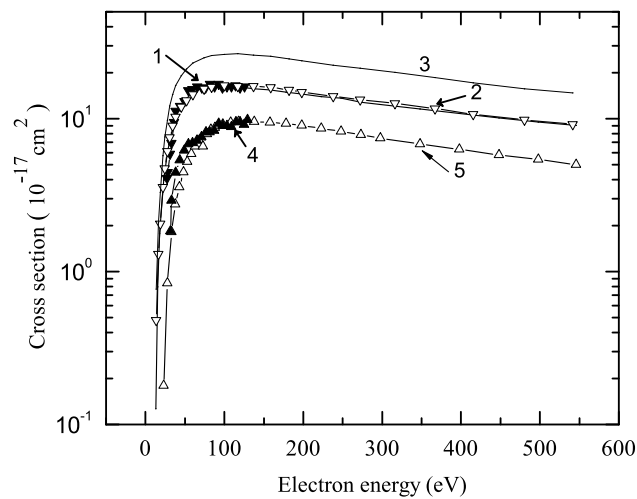}
\caption{(Color online) Dependence of cross sections for production of O$%
_{2}^{+}$ and (O$^{+}+$ O$_{2}^{2+}$) ions on the electron energy in $e+$O$%
_{2}$ collision. Curves: 1 - present data for O$_{2}^{+}$ ions; 2 - data for
O$_{2}^{+}$ from Ref. \protect\cite{46}; 3 - data for O$_{2}^{+}$ ions from
Ref. \protect\cite{47}. Curves: 4 - present data for (O$^{+}+$ O$_{2}^{2+}$)
ions; 5 - data for (O$^{+}+$ O$_{2}^{2+}$) ions from Ref. \protect\cite{46}.}
\label{Fig4}
\end{figure}

The probability of quenching as a function of target gas pressure in the ion source
for O$^{+}-$ Ar collisions was checked in Ref. \cite{30RL}. Typical emission spectra for the spectral line of Ar atom ($\lambda =731.1$
nm) are presented in Fig. \ref{Fig9}.
As it is seen, by increasing the O$_{2}$ gas pressure in the ion source
about 6 times (from 1.5$\times$10$^{-2}$ to 1.0$\times$10$^{-1}$ Torr) the
intensity of this line decreases by a factor of 1.4. This result indicates
that the presence of metastable ions plays a definite role in excitation of
Ar($6s$) state) \cite{30RL}.
The results for molecular oxygen ionization, O$_{2}^{+}$, production of
fragment ions, O$^{+}-$ O$_{2}^{2+}$ and O$^{2+}$, by electron impact are
investigated in Ref. \cite{46}. It is shown, that all processes have
threshold character, with broad maxima in the $100-200$ eV energy range.
Theoretical calculations for O$_{2}^{+}$ molecule ions are presented in Ref.
\cite{47}.
It should be noticed that in both of these works \cite%
{46,47} there is no information at all about the existence of metastable
particles in the beam. The data for the excitation function of total O$^{+}$
ions and metastable O$^{+}$ ions produced in $e+$O$_{2}$ collisions, from
threshold to 450 eV energy interval, are presented in Ref. \cite{48}. The
authors observed interesting but surprising results: the cross section for
the ion production in the ground \ and metastable states falls off more
slowly than metastable ions as the electron energy increases to
energies larger than 150 eV. Although production of metastable O$^{+}$ ions
was confirmed, no information regarding the influence of the pressure
condition in the ion source on the formation of metastable O$^{+}$ was
presented in Refs. \cite{46,47,48}. We address this situation by performing
measurements with variable electron energy and pressure condition in the
ion source using our approach.
Our
results for the energy dependence for electron impact ionization of
molecular oxygen, O$_{2}^{+}$, (curve 1 in Fig. \ref{Fig4})
as well as for production of fragment ions, O$^{+}-$ O$_{2}^{2+},$
(curve 2 in Fig. \ref{Fig4})
obtained in arbitrary units are normalized to the results from \cite{46}.
These data, along with the results for O$_{2}^{+}$ ions obtained in \cite{47}
are presented in Fig. \ref{Fig4}.

\begin{figure}[t]
\centering
\includegraphics[width=10.0cm]{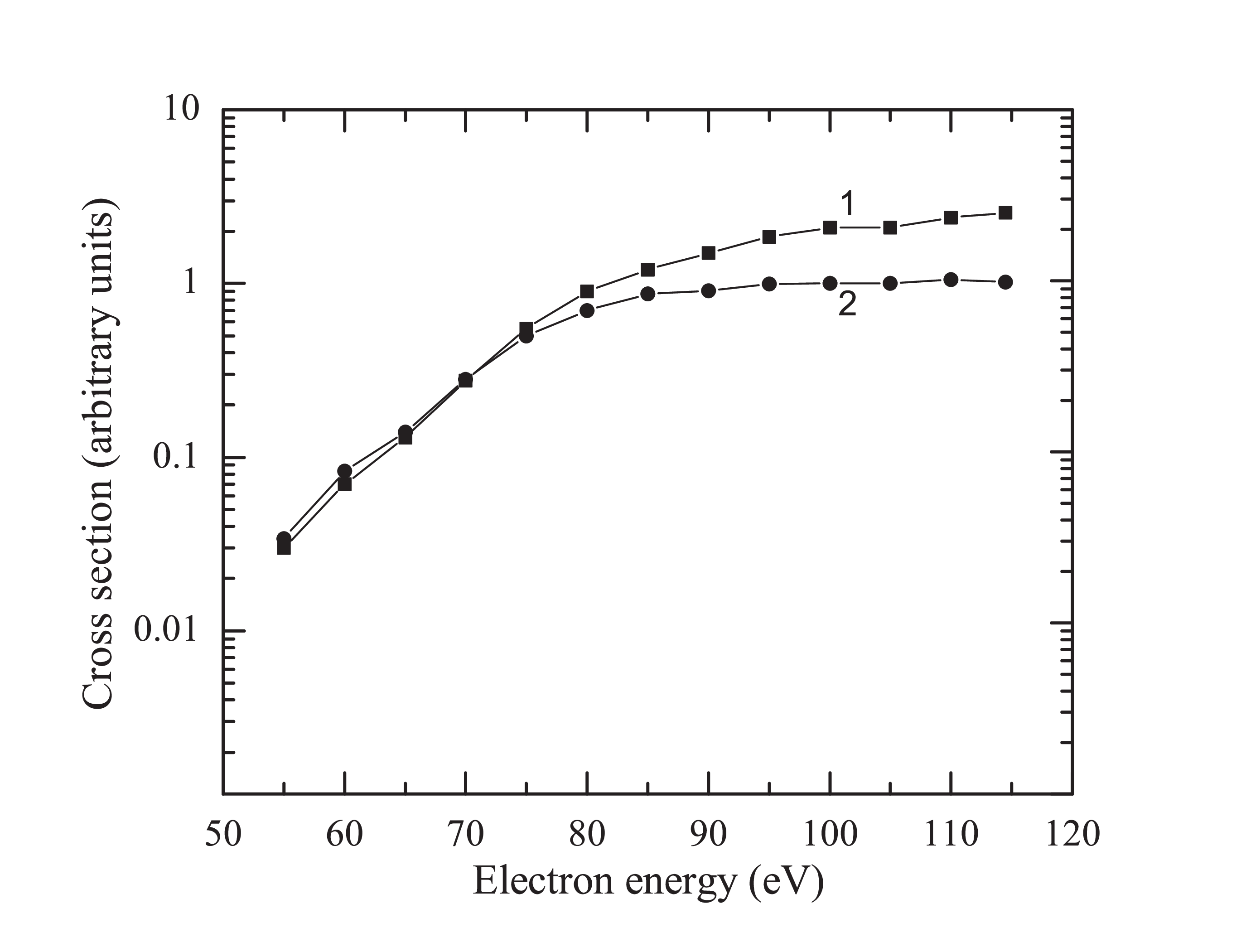}
\caption{(Color online) The cross section of O$^{+}+$O$_{2}^{2+}$ ions
production in $e+$O$_{2}$ collision as a function of the electron energy for
two different pressure conditions. Curves: 1 - data at the pressure $%
P=1\times10^{-4}$ Torr; 2 - data at the pressure $P=5\times10^{-4}$ Torr.}
\label{Fig5}
\end{figure}
It should be noted that the individual cross section for O$^{+}$ and O$%
_{2}^{2+}$ cannot be separately determined because each set of ions has the
same mass-to-charge ratio in the mass spectrometer. As to O$_{2}^{++}$
fragment ions, formed also in ionization processes, its value is by two
orders of magnitude less compared to other fragment ions and are not
presented in Fig. \ref{Fig4}. As it is seen from Fig. \ref{Fig4}, the
agreement between our results for O$_{2}^{+}$ (curve 1) and O$^{+}+$ O$%
_{2}^{2+}$ (curves 4) and the results from \cite{46} (curve 2) and (curve 5),
respectively, are excellent. A similarity of the energy dependence of the
cross section for O$_{2}^{+}$ is observed also with the theoretical results
(curve 3) reported in \cite{47} but it is not consistent with absolute value
of our results and with the results obtained in \cite{46}. For convenience
if we normalize the results of the relative cross section from \cite{47} to our
and to the results of \cite{46} (not shown in Fig. \ref{Fig4}), it will show
excellent agreement. All presented data are obtained under single collision
conditions ($3\times10^{-6}$ Torr). As it is well known (see \textit{e.g.}
\cite{32}), at such a pressure condition, the production of metastable
particles takes place, which can significantly affect the value of the cross
section. For this reason, for the $e-$O$_{2}$ collision system, we carried
out the measurements for production of ions (O$^{+}+$ O$_{2}^{2+}$) in the electron energy
range of $55-120$ eV and for two different pressure
condition: 1$\times$10$^{-4}$ Torr and 5$\times$10$^{-4}$ Torr. The results,
normalized with respect to each other, are given in arbitrary units in Fig. %
\ref{Fig5}. As it follows from Ref. \cite{32} the increase of working
pressure in the ion source is strongly related to an increase of a quenching
effect. Hence, the difference observed between the curves indicates the
presence of metastable particles in the ion beam. It means that a more
sophisticated exploration is needed by using an intense ion beam to control
the influence of metastable particles on the cross section. This issue can
be addressed by using a RF ion source \cite{30RL}.

Our ion source allows for varying and tuning the electron-impact ionization energy and the target-gas pressure. The large discrepancy between an abundance
of a metastable state can be attributed to different operating condition:
the pressure and electron energy of the ion source.
We can conclude that the confidence in the cross section is largely dependent on the
composition of the beam. To control the ion beam's metastable species, both
the precise exploration of the pressure condition in the ion source and evaluation
of the electron energy, is important.

The most suitable method, which allows evaluating the role of the influence of
metastable states is given in Ref. \cite{30RL} and is related to the usage of the radio frequency (RF) ion source.
One important advantage of the RF ion source is that, by changing the pressure in the ion source, the formation of a certain
fraction of metastable states can be found and identified easily.

The presence of ions in the metastable $^{2}P$ states was monitored by
measurement of the emission cross section of the first negative band system
(0,0) $\lambda =391.4$ nm of N$_{2}^{+}$ ions \cite{GKezLom2010} for different pressure
conditions in the RF source the energy of electrons used for the
production of O$^{+}$ ions remained  fixed. As a result, an influence of the pressure
condition in the RF source on the formation of certain O$^{+}$($^{2}P$)
states was well established.

In addition, the exact edge of the pressure condition in the RF ion source for $%
^{2}P$ state was revealed, which enabled us to distinguish ground state
and metastable state ions in the primary beam. The attempt to estimate
metastable $^{2}D$ state ions in the ion source was checked by measurement
of the band system of N$_{2}^{+}$, (3, 0) $\lambda =687.4$ nm and (4,
1) $\lambda =703.7$ nm for O$^{+}-$N$_{2}$ collision system \cite{30RL}.
Experiment showed that, despite the quasiresonant character of these
processes (energy defect = 0.06 eV) these bands are hardly excited,
which is probable due to the fact that either the probability for formation
of $^{2}D$ (at least for the energy range of 1-10 KeV) is low compare to O$%
^{+}$($^{2}P$) or/and O$^{+}$($^{4}S$)) or it strongly depends on a
peculiarity of the mechanism of formation of these ions formed during the
dissociative ionization in the collision of electrons with oxygen molecules in
the RF ion source.

We conclude that to obtain reliable results on the cross sections for
inelastic processes and to determine the mechanism of the formation of O$%
^{+}$ ions it is not sufficient to control just the pressure in the ion source
because the role of the electron energy, for production of ion, is crucial
and should be taken into the consideration.
It is especially worth
mentioning that this goal was achieved by the electron impact ion source where
the ionization chamber and ion source are united in the same space. The latter allows for varying and tuning the electron-impact ionization energy and the target-gas pressure.

\section{CONCLUSIONS}

The apparatus and experimental method are presented that
permit direct measurements of absolute partial cross sections
for electron-impact ionization and fragmentation of molecules. The apparatus is
simple in concept and embodies the ion source with a collision chamber and a mass spectrometer with a quadruple filter as a selector for
the product ions. The combination of these two devices
makes it possible to unequivocally collect all energetic fragment
ions formed in ionization and dissociative processes and to detect them with known efficiency.
It should be especially noted that we
used the ion source as the collision chamber. Our ion source allows to vary and tune the electron-impact ionization energy and the target-gas pressure.
The latter allowed us to
investigate collisions inside the ion source with the precise variation of electrons energy for the given target-gas pressure. This significantly raised the
luminosity, and simplified and accelerated the measurement procedure.
We demonstrated that obtaining reliable results for the cross sections for
inelastic processes and determining the mechanism of the formation of O$%
^{+}$ ions are crucial to control the pressure
and electron energy for production of ions in the ion source.
It is not sufficient to control just pressure in ion source
because the role of electron energy for production of ions and the composition of the O$^{+}$ beam is correlated with the pressure
and should be taken into consideration. A
comparison of our results with other experimental and theoretical data shows
good agreement and proves the validity of our approach.
These investigations allow us to find conditions for reliable beams of the ground O$^{+}$($^{4}S$) and metastable O$^{+}$($^{2}D$) and O$^{+}$($^{2}P$) ions to study the excitation processes induced by these ions in
collisions with N$_{2}$ molecules. The latter measurement we perform using the
optical spectroscopy method \cite{30RL}. The same method can be used to find conditions for obtaining the ground and metastable states of N$^{+}$ ions using molecular nitrogen as an input. The established conditions for the reliable beam of the ground and metastable states of N$^{+}$ ions that can be used to study the excitation processes induced by these ions in collisions with N$_{2}$ and O$_{2}$ molecules using
optical spectroscopy method.  The latter allows performing excitation function measurements for the O$^{+}-$N$_{2}$, O$^{+}-$O$_{2}$, N$^{+}-$N$_{2}$ and N$^{+}-$O$_{2}$
collision systems and to identify precisely different processes.

\end{document}